\documentclass[a4paper,fleqn,usenatbib]{mnras}

\usepackage{mathptmx}

\usepackage[T1]{fontenc}
\usepackage{ae,aecompl}

\usepackage{graphicx}	
\usepackage{amsmath}	
\usepackage{amssymb}	
\usepackage{BibDef}
\usepackage{geometry}
\usepackage{lipsum}

\newcommand\msun{\, \rm M_\odot}

\newcommand\mgal{{M_{\rm g}}}

\title[SC disruption by a BHB]{Star cluster disruption by a massive black hole binary}

\author[Bortolas, Mapelli \& Spera]{
Elisa Bortolas,$^{1,2}$\thanks{E-mail: elisa.bortolas@oapd.inaf.it}
Michela Mapelli,$^{1,3,4}$
Mario Spera$^{1,3,4}$
\\
$^{1}$INAF, Osservatorio Astronomico di Padova, Vicolo dell'Osservatorio 5, I--35122, Padova, Italy\\
$^{2}$Dipartimento di Fisica e Astronomia ``Galileo Galilei'', Universit\`a di Padova, Vicolo dell'Osservatorio 3, I--35122 Padova, Italy\\
$^{3}$INFN, Sezione di Milano-Bicocca, Piazza della Scienza 3, I--20126 Milano, Italy\\
$^{4}$Institut f\"ur Astro- und Teilchenphysik, Leopold Franzens Universit\"at Innsbruck, Technikerstra\ss e 25/8, A--6020 Innsbruck, Austria
}

\date{Accepted XXX. Received YYY; in original form ZZZ}

\pubyear{2016}

\begin{document}
\label{firstpage}
\pagerange{\pageref{firstpage}--\pageref{lastpage}}
\maketitle

\begin{abstract}
Massive black hole binaries (BHBs) are expected to form as the result of galaxy mergers; they shrink via dynamical friction and stellar scatterings, until gravitational waves (GWs) bring them to the final coalescence. It has been argued that BHBs may stall at a parsec scale and never enter the GW stage if stars are not continuously supplied to the BHB loss cone. 
Here we perform several $N$-body experiments to study the effect of a $8\times10^4\msun$ stellar cluster (SC) infalling on a parsec-scale BHB. We explore different orbital elements for the SC and we perform runs both with and without accounting for the influence of a rigid stellar cusp (modelled as a rigid Dehnen potential). We find that the semi-major axis of the BHB shrinks by  $\gtrsim{}10$ per cent if the SC is on a nearly radial orbit; the shrinking is more efficient when a Dehnen potential is included and the orbital plane of the SC coincides with that of the BHB.  
 In contrast, if the SC orbit has non-zero angular momentum, only few stars enter the BHB loss cone and the resulting BHB shrinking is negligible. Our results indicate that SC disruption might significantly contribute to the shrinking of a parsec-scale BHB only if the SC approaches the BHB on a nearly radial orbit. 
\end{abstract}




\section{Introduction}

There is compelling evidence that supermassive black holes (SMBHs) lie at the centre of galaxies since the earliest times \citep{haehneltrees,fan2003,jiang2007,jiang2008,willott2007,willott2010,mortlock2011,venemans2013,banados2014,wu2015}. According to the hierarchical paradigm \citep{WhiteRees1978}, present-day galaxies assemble through the merger of several progenitors, some of which possibly hosting an SMBH at their centre. Thus, SMBH binaries (BHBs) are expected to form as outcomes of galaxy mergers  \citep{Begelman1980}.

While current observational evidences of the existence of close BHBs are rather scanty \citep{komossa2003, rodriguez2006, Dotti2012,liu2014, runnoe2015,yan2015, li2016}, the spiral-in and coalescence of a  BHB is expected to be   an important source of gravitational waves (GWs, \citealt{ThorneBraginskii1976}) in the frequency range of the Pulsar Timing Array \citep{Hobbs2010,Babak2016} and of future space-borne GW detectors (e.g. LISA, \citealt{Amaro-Seoane2017}).  Observing GW emission from BHB mergers would then give us a crucial insight on the co-evolution of SMBHs and their host galaxies \citep{Volonteri2003,SesanaHaardt2004,Koushiappas2006, SesanaVolonteri2007,Tanaka2009, SesanaGair2011}, and would be a key test for the hierarchical paradigm.

Early theoretical and numerical studies on the evolution of BHBs highlighted the possible existence of a `final parsec problem' (\citealt{Milosav2003a}). During the merger of their host galaxies, the two SMBHs sink toward the centre of the common potential well by dynamical friction \citep{Chandrasekhar1943,Milosav2001}. When they are sufficiently close to form a binary, slingshot ejections of stars  further reduce their orbital separation \citep{Saslaw1974}. 
{ However, the binary shrinking may slow down considerably and even stop, if
the  loss cone (i.e. the region of the phase space harboring stars that can
interact with the BHB) has been emptied and cannot be refilled effectively.
This is expected to happen when the binary separation is of the order of 1
pc, thus the BHB may never enter the GW emission stage \citep{Begelman1980, Milosav2001,Yu2002,Milosav2003a,Makino2004}.}

Several mechanisms have been identified as possible solutions for the  final parsec problem. If the BHB evolves in a gas rich nucleus, gas drag can efficiently dissipate its  binding energy and the BHB may reach the GW emission stage within $\sim$100 Myr, regardless of the loss cone refilling \citep[see e.g.][]{Escala2004,Dotti2006,Goicovic2016}. Alternatively, a massive perturber, such as another massive black hole, a star cluster (SC),  a giant molecular cloud,   or the compact core of an infalling dwarf galaxy  might have contributed to refilling the loss cone \citep[see e.g.][]{Perets2008,Matsui2009}. The  Brownian motion of the BHB was also proposed as a significant driver of loss cone refilling \citep{Milosav2001,Chatterjee2003,Milosav2003}, but recent studies suggest that the wandering-induced shrinking of the BHB is not efficient if the merger remnant is composed of $\gtrsim10^6$ stars \citep{Bortolas2016}. 

Recently, a number of semi-analytical and numerical studies showed that the BHB stalling does not occur when the merger is simulated \emph{ab initio} \citep{Berczik2006,Preto2011, Khan2011, Khan2012, Khan12b, Gualandris2012,Khan2016}. The most likely reason is that the merger remnant is generally non spherical, and possibly rotates \citep[e.g.][]{Yu2002, Khan2013, Vasiliev2014, Vasiliev2015, Holley-Bockelmann2015,Gualandris2017}. In triaxial potentials the loss cone can be efficiently replenished at all times, thanks to the action of non spherical torques, and the BHB coalescence can be reached in  few Gyr at most \citep[][{ Bortolas et al., in preparation}]{Preto2011, Khan2011, Gualandris2012,Khan2016}. 

Here we  {  test whether an additional process may replenish the loss cone}: the infall of a SC onto the BHB. SCs that form in proximity of a galactic nucleus are expected to rapidly sink to the central parsec by dynamical friction \citep{Gerhard2001,mcmillan2003,portegies2003,kim2003,kim2004,gurkan2005,Fujii2008}. In fact, SC disruption has been proposed as one of the most promising mechanisms to form nuclear SCs {\citep{capuzzo1993,capuzzo2008,Arca-Sedda2015,Arca-Sedda2016}}, including that of the Milky Way \citep{antonini2012}.

Several young massive SCs (such as the Arches and the Quintuplet) lie in the nucleus of the Milky Way, which is a relatively quiescent galaxy (see \citealt{portegies2010} for a recent review). Young massive SCs are even more common at the centre of galaxy mergers, which are known to trigger bursts of star formation (see e.g. \citealt{sanders1988}). Since galaxy mergers are suspected to lead to the formation of both BHBs and young SCs, the dynamical-friction induced infall of a young SC onto a parsec-scale BHB should be a rather likely event.

Here we perform direct $N$-body simulations to study  the infall of a SC on a parsec-scale circular BHB. The paper is organized as follows: in Section~\ref{sec:methods} we describe the numerical methods and initial conditions of the simulations; in Section~\ref{sec:results} we present our results; our conclusions are drawn in Section~\ref{sec:sum}.

\begin{figure*}
	\includegraphics[width=\textwidth]{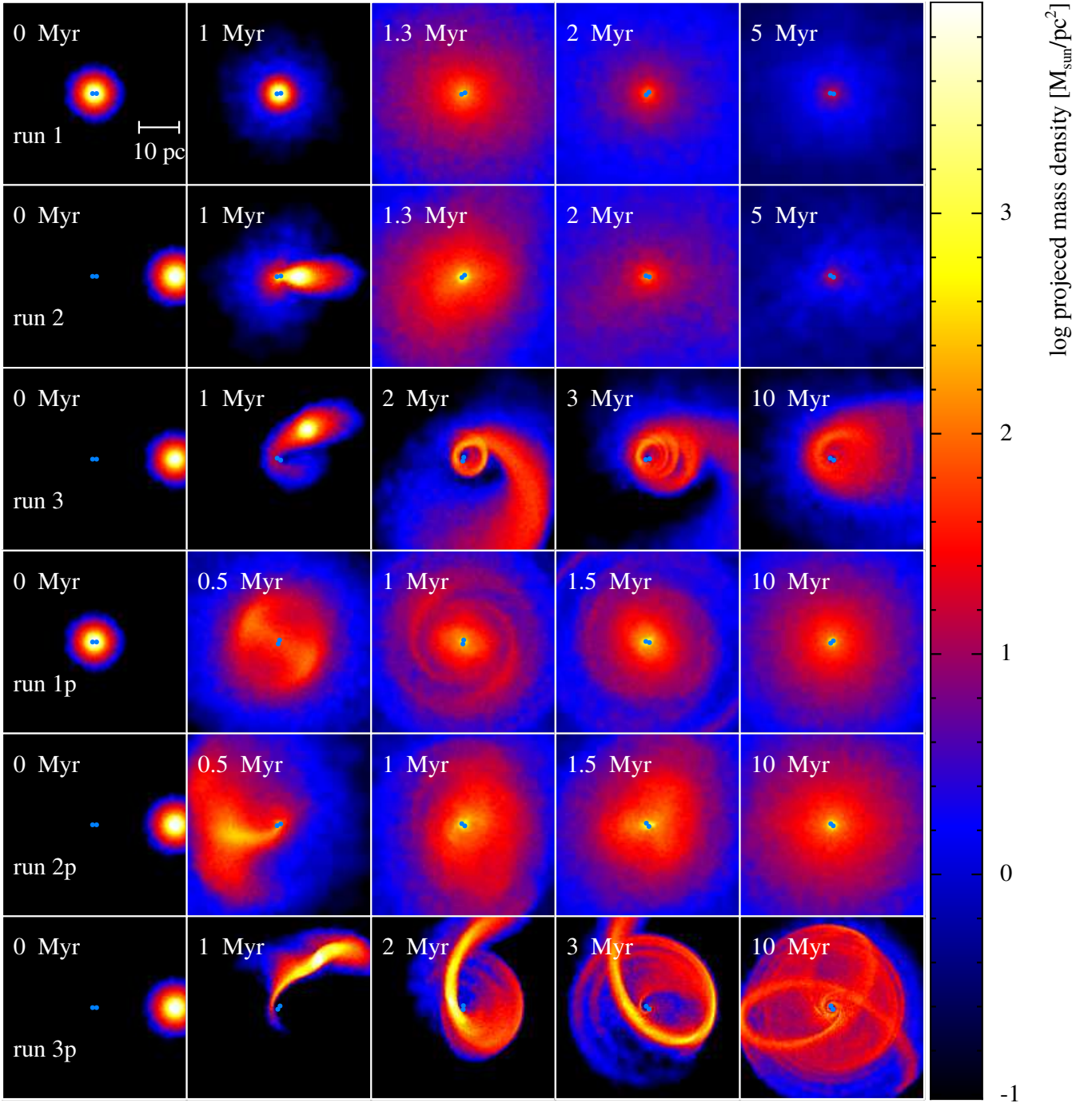} 
		\caption{ Time evolution of the stellar surface density projected on the $x-z$ plane -- i.e. the BHB orbital plane -- for runs~1, 2, 3 (top three rows), and runs 1p, 2p, 3p (bottom three rows). The blue central dots mark the position of the two SMBHs. The colour code refers to the smoothed projected mass density of stars, and the colour scaling is the same for all panels.}
    \label{fig:xz}
\end{figure*}

\begin{figure*}
	\includegraphics[width=\textwidth]{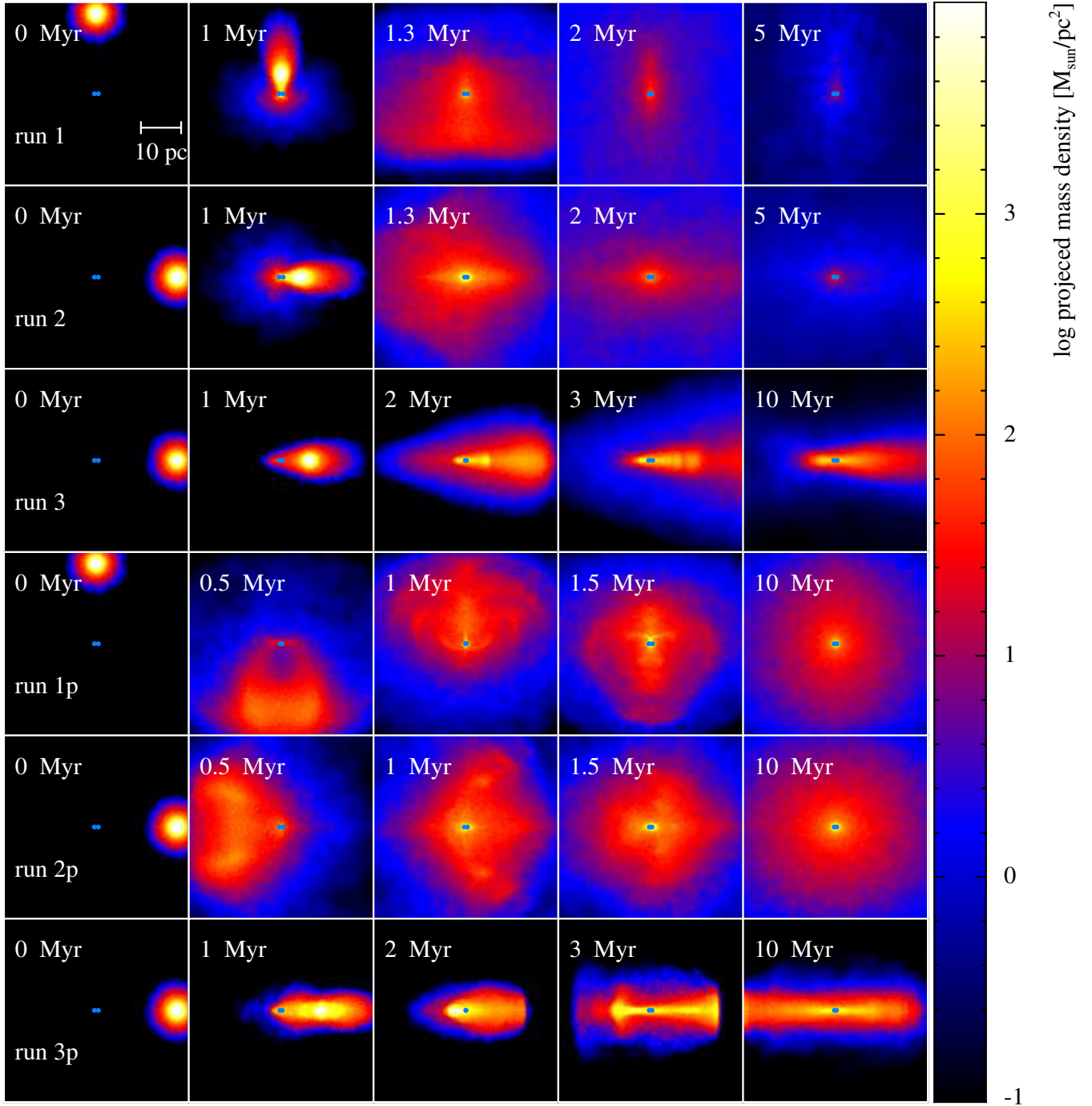} 
		\caption{Same as Fig.~\ref{fig:xz}, but the stellar surface density is projected on the $x-y$ plane.}
    \label{fig:zy}
\end{figure*}

\section{Methods}\label{sec:methods}

\begin{table}
  \centering
  \caption{Main features of the runs.}
  \label{tab:run}
  \begin{tabular}{lccc} 
    \hline
       Run & Galactic Potential &  Ang. Mom. & Orbit \\
     \hline
       run~1  & No  & No & Perpendicular \\
       run~2  & No  & No & Coplanar \\
       run~3  & No  & Yes& Coplanar \\
       run~1p & Yes & No & Perpendicular \\
       run~2p & Yes & No & Coplanar \\
       run~3p & Yes & Yes& Coplanar\\
    \hline
\end{tabular}
\begin{flushleft}
{ \footnotesize
For each run (Column~1) we report whether an underlying galactic potential is included (Column~2), if the initial orbit has some angular momentum (Column~3) and whether the orbit is coplanar or perpendicular with respect to the BHB orbital plane (Column~3).}
\end{flushleft}
\end{table}

In this paper we performed direct $N$-body simulations of the infall of a SC onto a parsec-scale BHB. To run the simulations we use the direct summation $N$-body code \textsc{HiGPUs} \citep{dolcetta2013}. \textsc{HiGPUs} implements a Hermite sixth order integration algorithm \citep{nitadori2008} with block time steps \citep{aarseth2003} and  has been designed to run natively on Graphics Processing Units.

To model the SC, we adopt a spherical King model \citep{king1966}, with central dimensionless potential $W_0=5$ and King's core radius $r_k=0.4$ pc. The SC is composed of $N=131070$ stars with masses distributed according to a \cite{kroupa2001} initial mass function,  with  mass range between $0.1 \,{}\textrm{M}_{\sun{}}$ and $100 \,{}\textrm{M}_{\sun{}}$. The initial total mass of the SC is $M_{\rm SC}\approx 8\times 10^4 \textrm{M}_{\sun{}}$. Stellar evolution is not included in the simulations.

Two SMBHs are placed in circular orbit in the $x-z$ plane with their centre of mass at the origin of the reference frame  and with angular momentum in the positive $y$ direction. The initial distance between the SMBHs is $1$ pc and each SMBH has mass $10^6$ $\textrm{M}_{\sun{}}$. 
 The softening parameter of the simulation is set to $\epsilon = 10^{-4}$ pc, which is several orders of magnitude smaller than the minimum distance reached by the SMBHs.
 
In this work we explored three different orbits for the cluster infall. For each of them we perform two runs: with and without including the underlying galactic potential. Namely, the potential of the host galaxy is included in runs 1p, 2p and 3p, while it is absent in runs 1, 2 and 3. When present, the galactic potential is described as a rigid potential, represented by a Dehnen's density profile \citep{Dehnen1993}:
\begin{equation}
\rho(r)=\frac{(3-\gamma) \,\mgal}{4\pi}\frac{r_0}{r^\gamma(r+r_0)^{4-\gamma}},
\end{equation}
 with total mass $\mgal=5\times10^{10} \msun$, scale radius $r_0=250$ pc and inner density slope $\gamma=0.5$. 

In runs 1, 1p, 2, and 2p the SC is initially in free fall, i.e. on a radial orbit. In runs 1 and 1p (runs 2 and 2p), the orbital plane of the SC is perpendicular (coplanar) with respect to the BHB orbital plane. Finally, in runs 3 and 3p, the SC is placed at the apoapsis of an eccentric orbit ($e=0.75$, defined through the periapsis and apoapsis distance in run 3p) with angular momentum along the $y$ axis, but with opposite sign with respect to the BHB angular momentum; this  maximises the relative velocity between the SC and the BHB. The centre of mass of the SC is initially located in $y = 20$ pc (runs 1 and 1p) and $x = 20$ pc (runs 2, 2p, 3 and 3p).  Runs~1 and~2 (runs~1p and 2p) are evolved for 5 Myr (10 Myr), while runs with angular momentum (3, 3p) are evolved for 20 Myr.  Table~\ref{tab:run} is a summary of the initial conditions of the six runs.

During the simulation, the centre of mass of the BHB is anchored to its initial position. To ensure this, we modified HiGPUs so that, after each time step, the binary centre of mass is re-centered at the origin of the reference frame and its velocity is set to zero. The BHB
recentering minimizes the binary wandering, which otherwise
would be too high in runs without the underlying galactic potential \citep[see ][]{Bortolas2016}. For consistency, we anchored the BHB centre of mass even in the runs including the galactic potential. We checked
that this choice does not affect the results of our simulations by re-running run~2p without the BHB anchorage. We find no appreciable differences in the evolution of the BHB and of the disrupted SC.

\section{Results}\label{sec:results}

{ Figures~\ref{fig:xz} and \ref{fig:zy} show the time evolution of the simulations in the $x-z$ plane (i.e. the BHB orbital plane) and in the $x-y$ plane, respectively. From these Figures, it is apparent that  the evolution of the system strongly depends on the initial angular momentum of the SC and on the presence of the Dehnen potential.}

\subsection{The evolution of the BHB without Dehnen potential}
In this section, we discuss the evolution of the BHB in runs~1, 2 and 3, in which we do not include a rigid Dehnen potential.

\begin{figure}
	\includegraphics[trim={10cm 0cm 2cm 16cm},angle=270,width=\columnwidth]{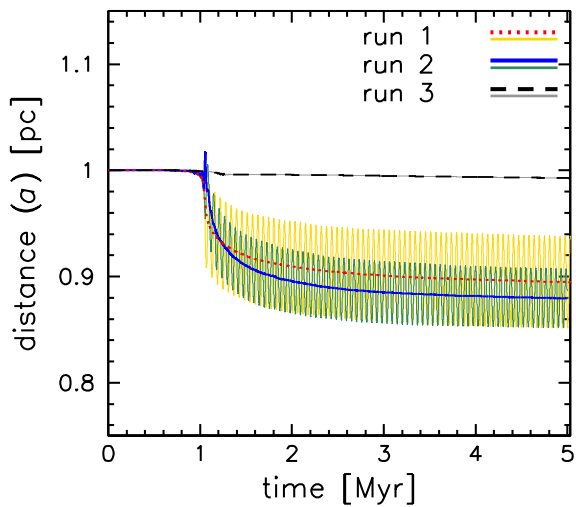} 
\caption{Evolution of BHB separation and semi-major axis as a function of time for the runs without Dehnen potential. Red dotted thick line (orange solid thin line): semi-major axis (separation) in run~1. Dark blue solid thick line (sea-green solid thin line): semi-major axis (separation) in run~2. Black dashed thick line (grey solid thin line): semi-major axis (separation) in run~3.}
	\label{fig:fig2}
\end{figure}

In both run~1 and 2, the SC starts interacting with the BHB at time $t\sim{}1$ Myr. During the interaction, stars belonging to the SC undergo three-body interactions with the BHB. Figure~\ref{fig:fig2} shows the evolution of the BHB orbital separation as a function of time. The orbital separation changes very fast during the first interaction with the SC, at $t\sim{}1-1.1$ Myr. Afterwards, the BHB keeps shrinking with a much shallower asymptotic trend,  and the change in the semi-major axis $a$ between 3.5 and 5 Myr is only  ${\rm d}a/{\rm d}t\sim{}-0.0025$ pc Myr$^{-1}$ in both runs.  We also note that the BHB keeps orbiting in the initial orbital plane. 

\begin{figure}
	\includegraphics[trim={10cm 0cm 2cm 16cm},angle=270,width=\columnwidth]{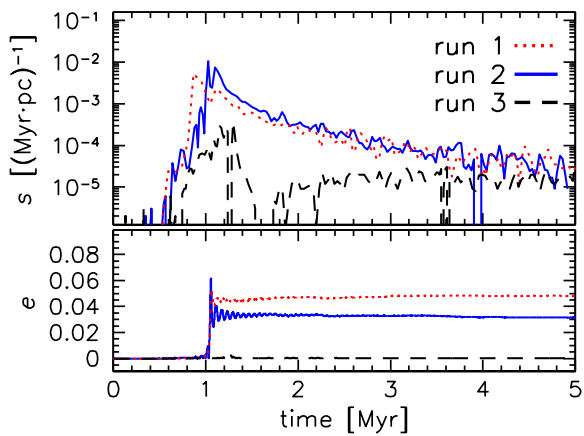} 
\caption{Evolution of the hardening rate ($s$, top panel) and of the eccentricity ($e$, bottom panel) of the BHB as a function of time for the runs without Dehnen potential. In all panels, red dotted line: run~1, blue solid line: run~2; black dashed line: run 3.
	\label{fig:fig3}}
\end{figure}

When the interaction is almost over (in less than 5 Myr), the BHB separation shrinks by $\sim 10$ and by $\sim{}12$ per cent in run~1 and run~2, respectively. This implies that the SC infall effectively replenished the loss cone of the BHB. As expected, the shrinking effect is more important in run~2, where the orbit of the SC and that of the BHB are coplanar. The reason is that the average relative velocity between the SMBHs and the stars is lower in run~2 than in run~1, maximizing the energetic exchange during the interaction. However, the difference of the final BHB semi-major axis between run~1 and 2 is only $\sim{}20$ per cent.

Similarly, Figure~\ref{fig:fig3} shows the behaviour of the hardening rate $s$, defined as
\begin{equation}\label{eq:eq1}
s(t)=\frac{{\rm d}}{{\rm d}t}\frac{1}{a},
\end{equation}
where $a$ is the semi-major axis of the binary. The hardening rate $s(t)$ quantifies the time variation of the BHB binding energy (the SMBH masses do not change with time).   $s(t)$ is maximum ($\sim{}10^{-2}$ Myr$^{-1}$ pc$^{-1}$)  during the first encounter between the SC and the BHB, and then it rapidly drops to few$\times\sim{}10^{-5}$  Myr$^{-1}$ pc$^{-1}$. The SC infall also produces a small but sudden jump in the BHB eccentricity: $e$ rises from 0 to $\sim{}0.05$ and to $\sim{}0.03$ after the first encounter with the SC in run~1 and run~2, respectively.

{ Figures \ref{fig:fig2} and \ref{fig:fig3} also show the time evolution  of the BHB in run~3 (with non-zero orbital angular momentum). The SC starts interacting with the BHB at $t\sim 1.2 - 1.3$ Myr, but the 
interaction is noticeably weaker with respect to runs~1 and 2. The BHB immediately shrinks of about the 0.4 per cent, while after 5 Myr its semi-major axis is only 0.8 per cent smaller than its initial value; even after 20 Myr the BHB  has shrunk by less than 1.5 per cent.   
In this run, the change in the BHB eccentricity is also negligible.  


\begin{figure}
	\includegraphics[trim={1cm 0cm 7cm 18cm},width=\columnwidth]{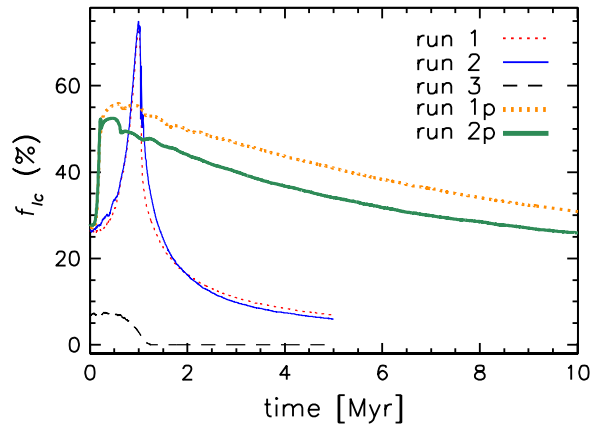} 
\caption{ Fraction of stars inside the loss cone (i.e. obeying to the condition in Eq. \eqref{eq:lc}) as a function of time. The plot shows the loss cone population in run~1 (thin red dotted line), run~2 (blue thin solid line), run 3 (black thin dashed line), run~1p (orange thick dotted line) and run~2p (sea-green thick solid line). Run~3p is not shown because the fraction of loss cone stars in run~3p is always below 0.3 per cent.
\label{fig:lc}}
\end{figure}


The difference between runs~1, 2 and 3 is related to the number of SC stars that are in the loss cone, defined as the region of the phase space harbouring stars with angular momentum per unit mass $j$ lower than the angular momentum per unit mass of  a circular binary with the same semimajor axis, i.e.
\begin{equation} \label{eq:lc}
j<J_{\rm LC}=\sqrt{2GM_{\rm BHB}a},
\end{equation}
  where $M_{\rm BHB}$ is the total mass of the BHB. The number of stars inside the loss cone\footnote{ Stars with positive energy (escapers) are not part of the loss cone population even if  their angular momentum is generally very low, as they will never interact again with the BHB.}  is shown in Figure \ref{fig:lc}.

In runs 1 and 2, about 27 per cent of the stars populate the loss cone at the beginning of the simulation. During the first approach, the SC is progressively stripped by the tidal forces of the BHB (see Figures~\ref{fig:xz} and \ref{fig:zy}). The tidal stripping forces many other stars to move on more radial orbits \citep{hills1991,perets2009} and funnels them inside the loss cone: at the maximum approach between the SC and the BHB, $\sim{}75$ per cent of the stars lie in the loss cone. The BHB expels stars very efficiently, and the loss cone population gets gradually depleted as the stars are scattered on highly energetic orbits: after 5 Myr, only $\approx 6$ per cent of the stars still inhabit the loss cone and they will likely become unbound in the next few Myr. 
  
The evolution of the loss cone population is totally different in run 3. The SC orbit in run~3 has non-zero angular momentum, thus stars satisfying the condition given in \eqref{eq:lc} are  initially only $\approx7$ per cent; in fact, the average angular momentum per unit mass of stars at the beginning of the simulation in run~3  is about twice the same quantity in runs~1 and 2.  In run 3, the fraction of stars in the loss cone is almost constant for the first Myr, because the non-zero angular momentum protects the cluster against the BHB-induced tidal stripping. When the SC reaches the maximum approach with respect to the BHB, the slingshot interactions between stars and BHB expel nearly all stars from the  loss cone, which is almost completely depleted.  

\begin{figure}
	\includegraphics[trim={0cm 0cm 6cm 15cm},width=\columnwidth]{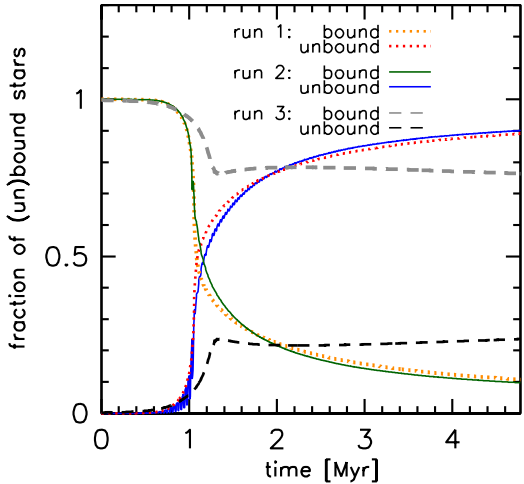} 
        \caption{ Fraction of stars bound or unbound to the BHB as a function of time for runs without Dehnen potential. At the beginning of the simulation all stars are bound to the BHB in run 1 (orange dotted line), run 2  (sea-green solid line) and run 3 (grey dashed line). When the BHB-SC interaction takes place ($t\sim 1$ Myr), the fraction of unbound stars (red dotted line for run 1, blue solid line for run 2, black dashed line for run 3) rapidly increases, because a lot of stars are ejected due  to the slingshot mechanism. 
    \label{fig:fig4}}
\end{figure}

Figure~\ref{fig:fig4} shows the fraction of stars bound and unbound to  the BHB as a function of time for runs without the Dehnen potential. In runs 1 and 2, the fraction of bound stars drops after the first interaction with the BHB. Only $10$ per cent of stars are bound to the BHB at the end of the simulation, regardless of the initial inclination between the SC and the orbital plane of the BHB. This implies that most stars in runs~1 and 2 receive a slingshot kick after the interaction with the BHB, sufficiently strong to unbind them from the BHB.  

In contrast, only $\sim 23$ per cent of stars escape the BHB potential in run~3. Given that the loss cone is almost empty after 5 Myr, stars still orbiting the BHB will probably not undergo slingshot ejections, unless stellar torques generated by other perturbers funnel such stars in the loss cone region. 

We further stress that the large fraction of unbound stars in runs~1 and 2 is a consequence of the absence of an underlying galactic potential in these runs. 



\subsection{The evolution of the BHB with a Dehnen potential}
In this section, we discuss the evolution of the BHB in runs~1p, 2p and 3p, which include a rigid Dehnen potential.

In runs~1p and~2p the first encounter between the SC and the BHB happens at $\sim{} 0.2$ Myr (Figs.~\ref{fig:xz} and \ref{fig:zy}) and the velocity of the SC at maximum approach is about twice that in runs~1 and 2, because of the Dehnen potential. Given the higher orbital speed, the SC in runs~1p and 2p is not entirely captured by the BHB during the first periapsis passage: the partially stripped remnant of the SC reaches the apoapsis and then falls back again onto the BHB  (while the SC is entirely stripped and captured by the BHB at the first periapsis passage in runs~1 and 2). The Dehnen potential keeps most stars bound to the system, so that they fall back to the centre and interact with the BHB several times. In contrast, most stars interact with the BHB only once and then are ejected from the system in the runs without Dehnen potential.


\begin{figure}
	\includegraphics[trim={10cm 0cm 2cm 16cm},angle=270,width=\columnwidth]{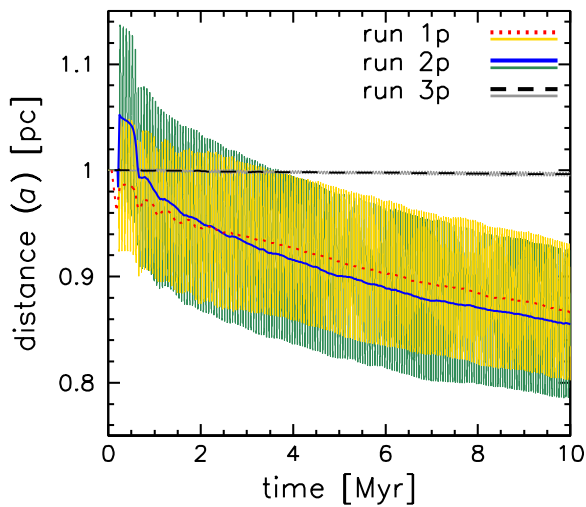} 
\caption{Evolution of BHB separation and semi-major axis as a function of time for runs with the Dehnen potential. Red dotted thick line (orange solid thin line): semi-major axis (separation) in run~1p. Dark blue solid thick line (sea-green solid thin line): semi-major axis (separation) in run~2p. Black dashed thick line (grey solid thin line): semi-major axis (separation) in run~3p.}
	\label{fig:fig2TF}
\end{figure}
\begin{figure}
	\includegraphics[trim={10cm 0cm 2cm 16cm},angle=270,width=\columnwidth]{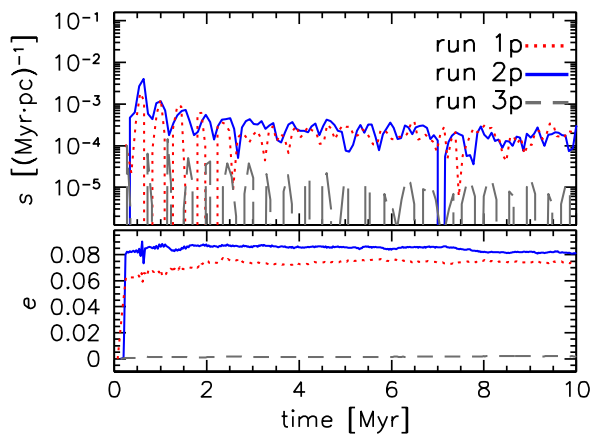} 
\caption{ Evolution of the hardening rate ($s$, top panel) and of the eccentricity ($e$, bottom panel) of the BHB as a function of time for runs with the Dehnen potential. In all panels, red dotted line: run~1p, blue solid line: run~2p; grey dashed line: run~3p.
	\label{fig:fig3TF}}
\end{figure}

 Figure~\ref{fig:fig2TF} shows the evolution of the BHB semi-major axis as a function of time for runs including the Dehnen potential.  
The final semi-major axis of the BHB in runs~1p and 2p is similar to what we found for runs~1 and 2: after 10 Myr, the BHB has shrunk by $\approx13$ per cent in run 1p and by $\approx 15$ per cent in run 2p. As we already discussed for runs~1 and 2, the BHB shrinking is slightly more efficient when the SC orbit is coplanar, because the relative velocity between the SC members and the BHB is lower.

However, there are several significant differences with respect to runs~1 and 2. 
In the first $\sim{}2$ Myr, the semi-major axis of the BHB does not shrink monotonically: it seems to undergo damped oscillations between smaller and larger values in both runs~1p and 2p (Figure~\ref{fig:fig2TF}). In run~2p the BHB semi-major axis even jumps to a value of $\sim{}1.05$ pc after the first interaction%
\footnote{As a matter of fact, a small jump in the binary semi-major axis also appears at the beginning of the SC-BHB interaction in run~2 (Fig.~\ref{fig:fig2}); however, the immediate  disruption of the SC by the binary hinders any further semi-major axis oscillation.}. 
%
%

This happens because, if the Dehnen potential is included, the BHB is a marginally soft binary (i.e. its binding energy is comparable to the average kinetic energy of an intruder, according to \citealt{heggie1975}) with respect to the SC as a single bullet. In fact, the binding energy of the BHB ($E_{\rm BHB}=G\,{}M_{\rm BH1}\,{}M_{\rm BH2}/2\,{}a\simeq{}4.3\times{}10^{52}$ erg, where $M_{\rm BH1}$ and $M_{\rm BH2}$ are the masses of the two SMBHs) is of the same order of magnitude as  the total kinetic energy of the SC ($E_{\rm K, SC}\simeq{}3.6\times{}10^{52}$) erg at the first periapsis passage ($E_{\rm K, SC}\approx 0.8 E_{\rm BHB}$ for runs~1p and 2p). While a hard binary (i.e. with binding energy much larger than the average kinetic energy of an intruder) tends to shrink after a gravitational encounter, a marginally soft binary might either increase or decrease its semi-major axis \citep{heggie1975}.

At the first periapsis passage the SC is still sufficiently compact to interact with the BHB as a single intruder, rather than a tidally disrupted system.  Thus, initially the BHB is rather soft with respect to the intruder and its semi-major axis tends to oscillate between larger and smaller values. At later times ($>2$ Myr), when the SC is disrupted, the BHB interacts with single stars (rather than with the SC as a whole). In this late stage, the BHB starts shrinking monotonically, because it is a hard binary with respect to each single star it interacts with.

The second important difference with respect to runs~1 and 2 is the derivative of the semi-major axis with time (${\rm d}a/{\rm d}t$). While in runs~1 and 2 the semi-major axis stalls after the first encounter (because the loss cone gets depleted), in runs~1p and 2p the semi-major axis keeps shrinking during the entire simulation. This is due to the fact that all stars in runs~1p and 2p remain bound to the system under the effect of the global potential. Thus, they can interact with the BHB more than once, when reaching the periapsis of their orbit.


This effect is also apparent from the hardening rate (Figure~\ref{fig:fig3TF}). The BHB hardening rate oscillates as a consequence of the oscillations in the BHB semi-major axis. 
The hardening rate  between 3.5 and 5 Myr is  $s\approx0.019$ ($s\approx0.014$) pc$^{-1}$ Myr$^{-1}$ in run~1p (run~2p), and even at later times (between 8.5 and 10 Myr) its value is of the order of $10^{-2}$ pc$^{-1}$ Myr$^{-1}$. Thus the BHB hardening efficiency in runs~1p and 2p is higher than in runs 1 and 2 at late times. 

From Figure~\ref{fig:fig3TF} it is also apparent that the eccentricity of the BHB  increases  almost instantaneously to $0.06-0.08$ during the first interaction, while it does not change significantly afterwards. This result is similar to what we find for runs~1 and 2.

The fraction of stars inside the BHB loss cone $f_{lc}$ for runs~1p and 2p is  shown in Figure~\ref{fig:lc}: at the beginning of the simulation, $f_{lc}$ is the same as in runs~1 and 2 ($\approx27$ per cent). The BHB-induced tidal stripping funnels about half of the SC members into the loss cone during the first $\approx 0.2$ Myr. The faster BHB-SC interaction in runs~1p and 2p results in a smaller number of stars initially funnelled in the loss cone; however, the depletion of the BHB loss cone is slower when the Dehnen potential is included. This results from the fact that stars may undergo repeated slingshot interactions before being definitely expelled from the loss cone. 

Figures~\ref{fig:fig2TF} and \ref{fig:fig3TF} also show the time evolution of the BHB in run~3p. Due to the initial orbital angular momentum of the SC, the fraction of stars in the loss cone is extremely low, and the binary shrinking is not effective. In particular, the BHB semi-major axis has shrunk by $\approx0.4\%$  after 10 Myr, when its hardening rate is only  $s\approx2.5\times10^{-4}$ pc$^{-1}$ Myr$^{-1}$, and even after 20 Myr it has shrunk only by $\approx0.6\%$. Initially no stars inhabit the loss cone in run~3p; 
after 10 Myr (20 Myr) only $\approx 0.16$ (0.27) per cent of stars are found in the loss cone. Also, the BHB eccentricity does not change significantly in run~3p (Figure~\ref{fig:fig3TF}). Thus, we can conclude that the properties of the BHB in runs~3 (without Dehnen potential) and 3p (with Dehnen potential) are very similar.}


\subsection{Structure of the SC remnant}
\begin{figure}
	\includegraphics[trim={0cm 0cm 6cm 15cm},width=\columnwidth]{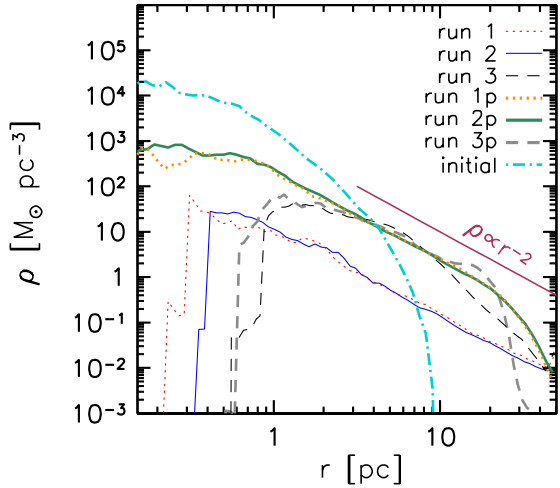} 
        \caption{Radial density distribution of stars  after 5 Myr (10 Myr) in runs 1, 2, 3 (1p, 2p, 3p). The final  density profile is compared with the initial density distribution (light-blue dash-dotted line). Thin dotted red line: run~1;  thin solid blue line: run~2;  thin dashed black line: run~3; thick dotted orange line: run~1p;  thick solid sea-green line: run~2p;  thick dashed grey line: run~3p. The initial density profile is computed with respect to the cluster center of mass, while the other density profiles are computed with respect to the BHB center of mass. 
    \label{fig:dens}}
\end{figure}
The interaction with the BHB leads to the complete disruption of the SC.  What is the final spatial distribution of stars around the BHB? Figures~\ref{fig:xz} and~\ref{fig:zy} show the projected stellar surface density in the $x-z$ and $x-y$ plane, while Figure~\ref{fig:dens} shows the final radial density distribution of stars. 

In runs~1 and 2, the disruptive interaction lowers the central density of the SC by $2-3$ orders of magnitude. Most of the SC mass is scattered out of the initial tidal radius ($r_{\rm t}=10$ pc) and the final density of the SC behaves as $\rho(r) \propto r^{-2}$. The stars keep memory of their initial orbit, because stars still bound to the BHB  after 5 Myr  distribute  on a prolate (triaxial) morphology in run~1 (run~2), whose longest axis lies along the cluster infall direction. 

In runs~1p and~2p the stellar distribution after 10 Myr is almost isotropic and less influenced by the initial conditions. The density distribution is cored within the BHB orbit, and decreases as $\rho \propto r^{-2}$ between 1 and 20 pc, while it rapidly vanishes  at larger distances. 
In fact, the density profile $\rho \propto r^{-2}$ resulting from the four radial runs is approximately what we expect for a relaxed stellar system around a point mass;  \citet{antonini2012} simulated the infall of several globular clusters on a SMBH and obtained a similar trend for the density at large radii.

If the SC is initially on a non-zero angular momentum orbit, 
 stars settle on a disc-like structure (aligned with the initial SC orbit), whose external radius is $R\lesssim{}20$ pc and  whose thickness is $\lesssim{}0.1\,{}R$. If the Dehnen potential is not included, the SC remnant is strongly asymmetric, as the BHB potential is almost Keplerian and most stars keep orbiting along the initial SC orbit. If the Dehnen cusp is present, the additional potential induces a precession on the stellar orbits, which results in a  three-lobed overdensity within the disc (bottom-right panel of Figure~\ref{fig:xz}). The  density profile of stars in runs~3 and 3p is also shown in Figure~\ref{fig:dens}, but one should keep in mind that the final stellar distribution in these two runs is far from being spatially isotropic.


\subsection{Distribution of the stellar orbital elements}

We now focus on the orbital parameters of stars that remain bound to the BHB and to the Dehnen potential (if present) by the end of the simulation.  We stress that this study is limited by the fact that we cannot take in account any possible dynamical interaction between SC stars and the pristine nuclear stellar population.

\begin{figure}    	
	\includegraphics[trim={0cm 0cm 0cm 7.5cm},width=.5\textwidth]{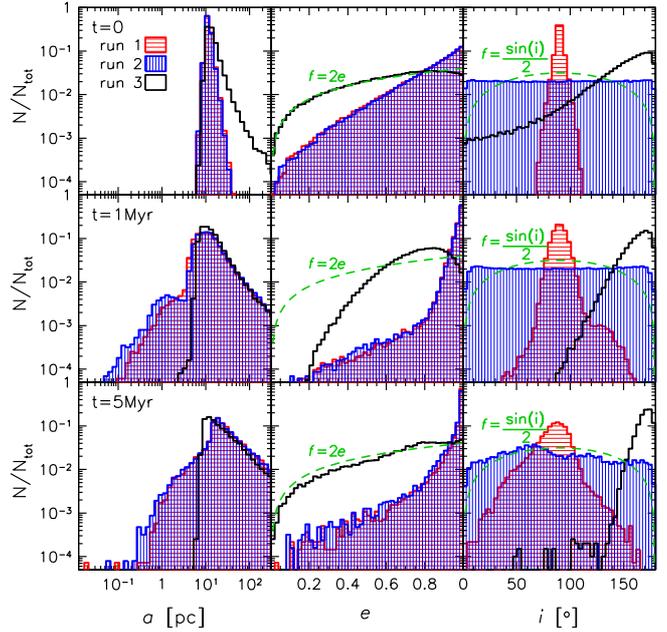} 
\caption{ Distribution of semi-major axis $a$ (left-hand panels), eccentricity $e$ (central panels) and inclination $i$ (right-hand panels) of the stellar orbits ($i$ is measured with respect to the plane of the BHB)  in run 1 (red, horizontally shaded histograms), run 2 (blue, vertically shaded histograms) and run 3 (black empty histograms) at different simulation times: the top row shows the initial distribution ($t=0$), the central row shows the distribution at $t=1$ Myr and the bottom row shows the distribution at $t=5$Myr. Green dashed line in the central panels: thermal eccentricity distribution $f(e)\ {\rm d}e=2e\ {\rm d} e$ \citep{Jeans1919}; green dashed line in the right-hand panels: isotropic distribution of inclinations $f(i)\ {\rm d} i = \sin(i) /2\  {\rm d} i$.\label{fig:fig5}}
\end{figure}

\begin{figure}    	
	\includegraphics[trim={0cm 0cm 0cm 7.5cm},width=.5\textwidth]{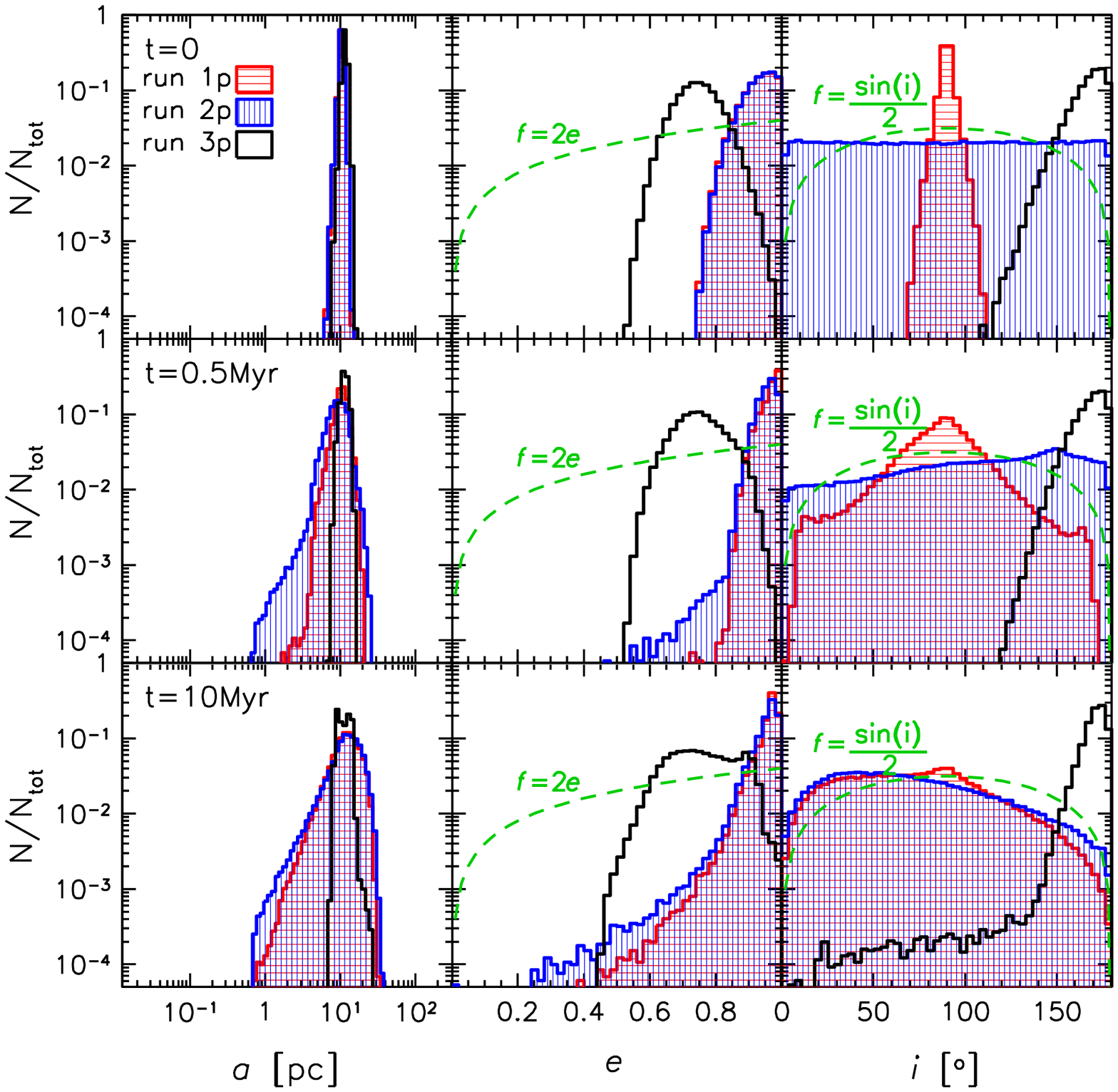} 
\caption{ Distribution of semi-major axis $a$ (left-hand panels), eccentricity $e$ (central panels) and inclination $i$ (right-hand panels) of the stellar orbits ($i$ is measured with respect to the plane of the BHB)  in run 1p (red, horizontally shaded histograms), run 2p (blue, vertically shaded histograms) and run 3p (black empty histograms) at different simulation times: the top row shows the initial distribution ($t=0$), the central row shows the distribution at $t=0.5$ Myr and the bottom row shows the distribution at $t=10$Myr. Green dashed line in the central panels: thermal eccentricity distribution $f(e)\ {\rm d}e=2e\ {\rm d} e$ \citep{Jeans1919}; green dashed line in the right-hand panels: isotropic distribution of inclinations $f(i)\ {\rm d} i = \sin(i) /2\  {\rm d} i$.\label{fig:fig5TF}}
\end{figure}

Figure~\ref{fig:fig5} shows the orbital properties of stars  bound to the BHB in runs~1, 2, and 3 (without Dehnen potential), while 
Figure~\ref{fig:fig5TF} shows the properties of stars bound to the BHB and the Dehnen potential for runs 1p, 2p, and 3p. The stellar orbital properties are shown  at different times: (i) at the beginning of the integration, (ii) when the interaction between the SC and the BHB just started, and (iii) after 5 Myr (10 Myr) in runs without (with) the Dehnen rigid potential.

\begin{table}
  \centering
  \caption{Fraction of bound stars with semi-major axis $a$ smaller than 1, 5, 10 and 20 pc.}
  \label{tab:a}
  \begin{tabular}{lrrrr} 
    \hline
       & $a$<1 pc & $a$<5 pc & $a$<10 pc & $a$<20 pc \\
     \hline
       run~1 & 0.25\% & 5.2\% & 12\% & 42\% \\
       run~2 & 0.64\% & 6.0\% & 13\% & 43\% \\
       run~3 & 0      & 0     & 8.2\% & 53\% \\
       run~1p & 0.03\% & 4.5\% & 32\% & 92\% \\
       run~2p & 0.14\% & 6.4\% & 31\% & 90\% \\
       run~3p & 0      & 0     & 36\% & 99.9\% \\       
    \hline
\end{tabular}
\begin{flushleft}
{\footnotesize
Column 1: run name; Columns $2-5$: fraction of bound stars whose semi-major axis is $<1$ pc (Column 2); $<5$ pc (Column 3); $<10$ pc (Column 4); $<20$ pc (Column 5). The listed fractions are computed after 5 Myr from the beginning of the simulation in runs 1, 2, and 3 and after 10 Myr in runs 1p, 2p, and 3p.}
\end{flushleft}
\end{table}

 In all runs, the distribution  of star semi-major axis initially peaks around $\sim{}10$ pc, which is the initial average semi-major axis of stars in the SC. 
 After the interaction, the distribution of semi-major axes becomes  noticeably broader, especially in runs~1 and~2. Table~\ref{tab:a} lists the fraction of non-escaping stars whose semi-major axis is smaller than a given threshold value; if the initial orbit is radial, $\lesssim 0.5$ per cent of stars have semi-major axis smaller than 1 pc: their final orbits are inside the semi-major axis of the BHB and they may further interact with it. While this percentage is small, these stars can have a further effect on the binary orbital shrinking. The fraction of stars with  semi-major axis smaller than 5 pc is higher when the radial infall is coplanar (runs 2 and 2p), as the coupling between the SC and the BHB is stronger and more stars settle on low-energy orbits tightly bound to the BHB. 


The distribution of inclinations $i$ of the stellar orbit  with respect to the BHB orbital plane strongly depends on the initial orbital plane of the SC in runs 1, 2 and 3: while bound stars in run~1 preserve a nearly perpendicular orbital inclination with respect to the orbital plane of the BHB, the distribution of bound stars in run~2 is more isotropic. In run~3, the SC-BHB interaction initially drives all stars with $i<90$ degrees on orbits with $i\gtrsim 90$ degrees; by the end of the run almost all stars in run~3 rotate in the opposite direction to the BHB,  except for a few stars that probably experienced a strong interaction with the BHB. The counter rotation of most stars 
results from the choice of giving to the SC an orbital angular momentum opposite to that of the BHB.

The final distribution of stellar inclinations in runs with the Dehnen potential is similar for runs 1p and 2p, suggesting that stellar inclinations may reach the same equilibrium configuration if one waits long enough; however, after 10 Myr a small fraction of stars in these runs still keep memory of the initial conditions of the simulation (e.g. there is a small peak in the distribution of inclinations at $\sim90$ degrees in run~1p). In addition, $\approx 70$ per cent of bound stars in runs~1p and 2p settle on orbits whose inclination is smaller than 90 degrees; this indicates that stars preferentially align their orbital angular momentum with the BHB one. Such result is not surprising as gravitational torques induced by the binary can drag a number of stars into corotating orbits \citep[e.g.][]{mapelli2005}.
Despite the fact that in run~3p most stars keep memory of their initial inclination ($\sim 180$ degrees), even in this run some stars are dragged onto corotating orbits by the gravitational torques of the binary, as 0.3 per cent of them have an inclination smaller than 90 degrees after 10 Myr.

In all radial runs 
the eccentricities of bound stars are much higher than a thermal distribution, as stars keep memory of the  initial  radial orbit of the SC; the eccentricity distributions look similar for such radial runs even if perpendicular runs 1, 1p always have a slight  overabundance of highly eccentric stars compared to the coplanar runs 2, 2p.
The large fraction of very eccentric objects also indicates that most bound stars are only marginally bound to the system. 

The distribution of eccentricities in run~3 is initially very similar to the thermal distribution \citep{Jeans1919}, but this is probably because the initial orbital eccentricity of the SC is $e=0.75$, i.e. it is really close to the mean-square value of the thermal eccentricity distribution. At later times, such distribution is still close to the thermal one, but more power is found at high eccentricities, indicating  that the BHB funnels stars on more radial orbits.

The initial eccentricity distribution of stars in run~3p still peaks at $e=0.75$, but the higher initial velocity of the SC in this run makes the distribution narrower compared to the one of run~3. At late times, the eccentricity distribution of run~3p slightly broadens and a small peak is found at $e\approx 0.9$, as a probable signature of weak slingshot interactions.

\subsection{Hyper-velocity stars}

\begin{figure}
	\includegraphics[trim={0cm 0cm 6cm 15cm},width=\columnwidth]{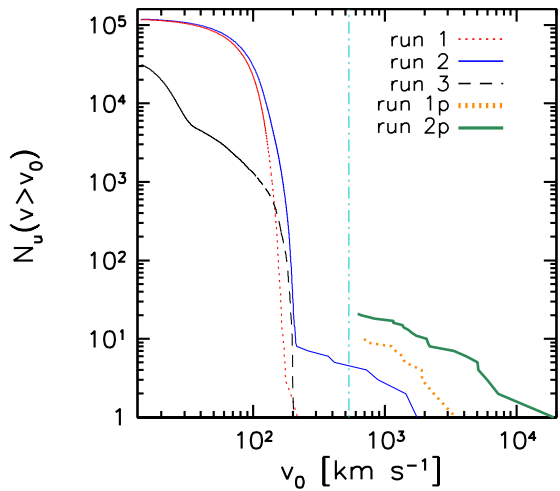} 
        \caption{Cumulative distribution of unbound stars $N_{\rm u}$ whose radial  velocity is greater than a threshold velocity $v_0$ as a function of  $v_0$ after 5 Myr in runs 1, 2 ,3 and after 10 Myr in runs 1p, 2p.  Red dotted thin line: run~1; blue solid thin line: run~2; black dashed thin line: run~3; orange dotted thick line: run~1p; sea-green solid thick line: run~2p; no stars escape from the system in run~3p. The dash-dotted vertical line at $v_0=533$ km s$^{-1}$ shows the escape velocity from the Milky Way  at three virial radii \citep{Smith2007,Piffl2014}.
    \label{fig:vel-esc}}
\end{figure}

Figure~\ref{fig:vel-esc} shows the cumulative distribution of the radial velocity $v_r$ of stellar escapers, i.e. stars that become unbound during the simulation;  here $v_r$ represents the component of the stellar velocity vector projected along the radial direction.  From  Figure~\ref{fig:vel-esc}, it is apparent that a large number of stars become unbound in runs~1, 2 and 3 (without Dehnen potential), but their velocity is $>100$ km s$^{-1}$ only in few cases.

The fastest escapers in runs~1 and 3 attain a velocity $v_r\sim{}200$ km s$^{-1}$ at most. Run~2 produces a marginally larger number of fast escapers: four objects attain an escape speed greater than the estimated escape speed from the Milky Way at three virial radii: $\approx533$ km s$^{-1}$ according to \citet{Piffl2014}.

In contrast, Figure~\ref{fig:vel-esc} shows  that only few tens of stars become unbound in runs~1p and 2p (with Dehnen potential), but their velocity is always $>600$ km s$^{-1}$. In particular, only 10 and 21 stars out of $10^5$ get unbound in run~1p and 2p, respectively, while no star leaves the potential well in run~3p. 

The velocity attained by the escapers in runs~1p and 2p can reach values as high as 5,000 km s$^{-1}$ and 20,000 km s$^{-1}$, respectively\footnote{If we do not anchor the BHB centre of mass to the origin of our reference frame, the number of produced hyper velocity stars is similar to what we show in Fig.~\ref{fig:vel-esc}; however, the maximum velocity attained by such hyper velocity stars drops to  about 6,000 km s$^{-1}$. }. Again, coplanar orbits (run~2p) are more efficient in producing  high velocity stars than perpendicular SC orbits (run~1p). We stress that all the escapers in runs~1p and 2p can be classified as genuine hyper-velocity stars  \citep{hills1988,brown2005,brown2006}, produced by the interaction with the BHB.

\section{Discussion and conclusions}
\label{sec:sum}
In this paper, we simulated the disruption of a SC by a BHB. We explored several different configurations for the SC-BHB interaction: with and without a rigid Dehnen potential, considering a SC orbital eccentricity $e=1$ and $e=0.75$, and assuming 0 or 90 DEG inclination between the orbital plane of the SC and that of the BHB. 

Runs without Dehnen potential are quite unrealistic, because the BHB is expected to lie in the centre of the galactic potential well. However, we ran them, because they represent a very simple test case, and because we can infer the role of the galactic potential from the comparison between runs with and runs without a galactic potential.

When the initial SC orbit has non-zero angular momentum ($e=0.75$), only few stars enter the BHB loss cone. As a consequence, the hardening rate of the BHB is almost negligible. This result is nearly unaffected by the presence of a Dehnen potential.

In contrast, if the initial orbit of the SC is radial ($e=1$), the infall of the SC effectively refills the loss cone of the BHB: the semi-major axis of the BHB changes by $\gtrsim$10 per cent within $5-10$ Myr.  
Even if nearly radial orbits are not expected to be common, they can be produced by collisions between molecular clouds. In particular, the collision of two molecular clouds close to the central parsec of a galaxy might trigger the formation of a SC with very low orbital angular momentum  (e.g. \citealt{Hobbs2009,Mapelli2012}). For instance, \citet{Tsuboi2015} recently showed that at least part of the star formation observed within the Galactic Centre may be triggered by collisions between molecular clouds.

In our simulations, if the Dehnen potential is not included and the SC infalls radially, the semi-major axis of the BHB shrinks very fast during the first encounter with the SC, but then it stalls. In contrast, if the Dehnen potential is included, the hardening of the BHB is initially less efficient, but then the BHB keeps shrinking at a significant rate for the entire simulation (10 Myr).
The reason of this difference is that most stars that interact with the BHB acquire a kick velocity of several ten to several hundred km s$^{-1}$. If only the SC and the BHB contribute to the potential well, these stars end up unbound and completely ejected from the system. Thus, they cannot undergo more than one interaction with the BHB. 

In contrast, if the rigid Dehnen potential is included, these kick velocities are too low to unbind a star: most stars remain bound to the potential well. After reaching their new apoapsis, these stars fall back toward the centre of the potential well and might interact with the BHB several times. Thus, the BHB keeps shrinking for a longer time, because each star can undergo multiple interactions with the BHB.

We stress that the simulations presented here neglect the effects  of dynamical friction induced by the pristine stellar population on the infalling SC (because the Dehnen potential is modelled as a rigid potential). Gravitational drag can bring the SC remnant on orbits closer to the BHB and further assist the BHB shrinking. However, \citet{Petts2017} recently showed that dynamical friction is inefficient in bringing SC stars on orbits closer than $\sim2$ pc from the central massive object(s), even if one assumes a very dense SC  infalling from $5-15$ pc distance.

Another effect we neglected is the loss cone refilling induced by a massive perturber (the SC, in our case) on the pristine stellar population, as this mechanism was already explored in a series of previous studies \citep{Perets2008,Matsui2009}: they demonstrated that a massive object is able to deflect the orbits of many stars belonging to the pristine galaxy core and funnel them onto the loss cone, enhancing the binary shrinking. In particular, \citet{Matsui2009} analysed how the BHB shrinking rate can be boosted by the infall of a compact dwarf galaxy merging with the BHB host galaxy.
They consider the infall of an object whose mass is $\sim 10$ times the BHB mass (while in our runs $M_{\rm SC}\approx 0.04 M_{\rm BHB}$) and they do not study the phase-space redistribution of stars belonging to the infalling stellar system, focusing on the effects of the dwarf-induced loss cone repopulation instead.

Finally, we stress that   in our simulations the change in the BHB semi-major axis was explored only for a limited number of cases, and in a forthcoming study we will investigate what happens for different orbital properties and masses of the SC and different BHB separations. However, we can already make some guess on the effect of a different choice of initial conditions, by means of some analytic consideration. The change of the semi-major axis likely depends on (i) the initial semi-major axis of the BHB, (ii) the initial relative velocity between SC and BHB, (iii)  the total mass of the SC, and (iv) the mass of the BHB.

 The geometric cross section of the BHB scales as the square of the semi-major axis. Thus, the effect of SC infall would have been stronger for a wider BHB, because all SC members would have passed inside the separation between the two SMBHs. However, this effect is mitigated by gravitational focusing: the trajectory of a SC star is deflected by the gravitational pull of the BHB. 
Accounting for gravitational focusing, the effective periapsis distance between the centre-of-mass of the SC and that of the BHB is $p\approx{}b^2\,{}v_{\rm i}^2/[2\,{}G\,{}(M_{\rm BHB}+M_{\rm SC})]$, where $b$ is the impact parameter, $v_{\rm i}$ is the initial relative velocity between the BHB and the SC, $G$ is the gravitational constant, $M_{\rm BHB}$ is the BHB mass, and $M_{\rm SC}$ is the SC mass \citep{sigurdsson1993}. In our runs~1, 2, 1p, and 2p, we chose $v_{\rm i}=0$, which implies $p\sim{}0$. Thus,  the result of these runs can be considered as an upper limit to the effect of three-body encounters on the BHB shrinking.
For a small impact parameter (of the same order of magnitude as the BHB semi-major axis), the shrinking of the binary is completely determined by the ratio between the SC mass and the BHB mass, because the average relative change of the BHB binding energy $\Delta{}E_{\rm b}/E_{\rm b}$ per encounter scales as \citep{hills1983,quinlan1996,colpi2003,mapelli2005}
\begin{equation}\label{eq:eq2}
\frac{\Delta{}E_{\rm b}}{E_{\rm b}}=\xi{}\,{}\frac{m_\ast}{M_{\rm BHB}},
\end{equation}
where $m_{\ast}$ is the average mass of a single star, and $\xi{}$ is a dimensionless factor ($\xi{}\sim{}1-10$ for small impact parameters, \citealt{hills1983}). Equation~\ref{eq:eq2} implies that the expected variation of the BHB semi-major axis due to SC infall is
\begin{equation}\label{eq:eq3}
1-\frac{a_{\rm f}}{a_{\rm i}}\sim{}0.1\,{}\left(\frac{\xi{}}{1}\right)\,{} \left(\frac{N_\ast}{10^5}\right)\,{}\left(\frac{m_\ast/M_{\rm BHB}}{10^{-6}}\right),
\end{equation}
where $a_{\rm f}$ and $a_{\rm i}$ are the final and initial BHB semi-major axis, respectively, while $N_{\ast}$ is the number of stars in the SC. 

The change of semi-major axis derived from this back-of-the-envelope calculation is remarkably similar to the value we obtained from our runs~1, 2, 1p, and 2p (i.e. the simulations where the SC is on a radial orbit). Thus, we might expect that a SC with $N_\ast{}\gtrsim{}10^6$ on a nearly radial orbit would have lead a $\sim{}10^6$ M$_\odot$ BHB close to the regime where the orbital decay by GW emission is efficient. However, it must be kept in mind that only a small fraction of SC members can efficiently interact with the BHB once the semi-major axis has dropped to $<<1$ pc. Dedicated simulations are needed to probe this extreme situation.  Moreover, to derive equation~\ref{eq:eq3}, we implicitly assumed that each star scatters with the BHB only once (as in the runs without Dehnen potential). As we have discussed in Section~\ref{sec:results}, this assumption gives us a lower limit to the efficiency of BHB hardening.

In this paper, we also investigated the fate of the SC. In all runs, the SC is almost completely disrupted by the interaction with the BHB. SCs infalling with  non-zero orbital angular momentum settle on a disc-like structure, whose morphology strongly depends on the initial SC eccentricity and on the presence of a Dehnen potential. No hyper-velocity stars are produced if the SC orbit has non-zero angular momentum.

If the SC is on a radial orbit and the Dehnen potential is not included, $\sim 95$ per cent of stars are kicked onto unbound orbits and only a small fraction of the initial SC keeps orbiting the BHB. These bound stars settle into a small subsystem whose shape is strongly influenced by the initial SC orbit; the final distribution of stars  follows a trend $\rho{}(r)\propto{}r^{-2}$.

If the SC infall is radial and the Dehnen potential is included, most stars remain bound to the global potential and are generally confined within $\sim50$ pc from the BHB. Only few tens of stars become unbound in this case, but their velocities are of the order of several thousand km s$^{-1}$. Thus, these are genuine hyper-velocity stars. These features represent the main observational imprints of a SC that was recently disrupted by a BHB.

\section*{Acknowledgements}
We thank the anonymous referee for their useful comments and suggestions.
We also thank  Massimo Dotti, Alessia Gualandris, James Petts, Alessandro Trani, Alessandro Ballone, Fabio Antonini, Manuel Arca-Sedda, Roberto Capuzzo Dolcetta, Peter Berczik, Rainer Spurzem and Long Wang for useful discussions and suggestions. 
We acknowledge the CINECA Award N. HP10CP8A4R and  HP10C8653N, 2016 for the availability of high performance computing resources and support. Part of the Numerical calculations have been made possible through a CINECA-INFN agreement, providing access to resources on GALILEO and MARCONI at CINECA. We acknowledge financial support from the Istituto Nazionale di Astrofisica (INAF) through a Cycle 31st PhD grant, from the Italian Ministry of Education, University and Research (MIUR) through grant FIRB 2012 RBFR12PM1F, from INAF through grant PRIN-2014-14, from the MERAC Foundation, from the Fondazione Ing. Aldo Gini and from INAF-Osservatorio Astronomico di Arcetri through the `Stefano Magini' Prize.

\bibliography{biblio} 

\bsp	
\label{lastpage}
\end{document}